\documentclass[11pt,reqno,a4paper]{amsart}
\usepackage{comment}
\usepackage{MnSymbol}              % See geometry.pdf to learn the layout options. There are lots.
\usepackage{graphicx}
\usepackage{color,bm}
\usepackage{xcolor}
\usepackage{hyperref}
\usepackage{tikz}
\usetikzlibrary{patterns}
\usepackage{caption}
\usepackage{subcaption}
\usepackage{mhequ}
\hypersetup{colorlinks=true}
\newcommand{\beqn}{\begin{equation}}
\newcommand{\eeqn}{\end{equation}}
\newcommand{\beqna}{\begin{eqnarray}}
\newcommand{\ee}{\end{eqnarray}}
\newcommand{\best}{\begin{eqnarray*}}
\newcommand{\ees}{\end{eqnarray*}}
\newcommand{\bal}{\begin{align}}
\newcommand{\eal}{\end{align}}

\newcommand{\bp}{\begin{pmatrix}}
\newcommand{\ep}{\end{pmatrix}}
\newcommand{\al}{\alpha}

\newcommand{\be}{\beta}
\newcommand{\ga}{\gamma}

\newcommand{\e}{\epsilon}

\newcommand{\ol}[1]{\bar{#1}}

\newcommand{\ul}[1]{\underset{\bar{}}{#1}}

\newcommand{\xa}{\bar{u}}
\newcommand{\xub}{\underline{u}}
\newcommand\Pa{Painlev\'e }

\newtheorem{Def}{Definition}
\newcommand\pxp{\mathbb{P}^1\times\mathbb{P}^1}
\newcommand{\PX}{{\rm Pic}\,X}

\usepackage{layout}
\begin{document}
%\layout

\title{A systematic approach to reductions of type-Q ABS equations}

\author{Mike Hay} 
\address{Dipartimento di Matematica e Fisica, Universit\`a degli Studi Roma Tre and INFN Sezione di Roma Tre}

\author{Phil Howes}
\address{School of Mathematics and Statistics, The University of Sydney}

\author{Nobutaka Nakazono}
\address{School of Mathematics and Statistics, The University of Sydney}

\author{Yang Shi}
\address{School of Mathematics and Statistics, The University of Sydney}

\begin{abstract}

We present a class of reductions of M\"obius type for the lattice equations known as Q1, Q2, and Q3 from the ABS list. The deautonomised form of one particular reduction of Q3 is shown to exist on the $A_1^{(1)}$ surface which belongs to the multiplicative type of rational surfaces in Sakai's classification of Painlev\'e systems. Using the growth of degrees of iterates, all other mappings that result from the class of reductions considered here are shown to be linearisable. Any possible linearisations are calculated explicitly by constructing a birational transformation defined by invariant curves in the blown up space of initial values for each reduction. 

\end{abstract}

\maketitle

\section{Introduction}\label{introQreds}

Integrable lattice equations and mappings are discrete analogues of the well known integrable partial and ordinary differential equations. The integrability of discrete systems has been the focus of vigorous activity recently, from which it has emerged that geometry plays a central role. Of primary interest are lattice equations and discrete Painlev\'e equations, where the geometry manifests itself in various ways. In \cite{abs:03}, a set of lattice equations was constructed and classified by a property known as ``three-dimensional consistency", this set of equations is known as the ``ABS list" and has received much attention. These equations on quad graphs can be considered discrete analogues of integrable partial differential equations in 1+1 dimensions. The discrete Painlev\'e equations are discrete analogues of the continuous Painlev\'e equations, nonlinear second order differential equations possessing the Painlev\'e property: all moveable singularities are poles. It has been shown by Sakai \cite{sak:01} that discrete Painlev\'e systems arise naturally in the actions of Weyl (reflection) groups of affine type. 

It is well known that solutions of integrable PDEs invariant under the action of Lie groups satisfy ODEs of Painelev\'e type \cite{ KdV-PII:79, NS:80, Bou:80, Dubrovin:99,SineG:79,mKdV-PII:76}. The solutions are called similarity reductions of these PDEs. It was first found \cite{NijP:91} that a discrete \Pa equation arises as a similarity reduction of the lattice KdV equation much in the same way continuous \Pa equations can be derived as scaling reductions of integrable PDEs. Although, the connection between symmetry invariant solution of integrable lattice equations and the recent classification of discrete \Pa equations \cite{sak:01} is not yet well understood, there have been much interest and intense work in this direction. In particular, the main approaches for obtaining discrete \Pa equations by reductions of integrable lattice equations have been periodic constraints \cite{GramaniP:91,Gramani:05,rcg:09}, nonlinear constraints derived by the infinite matrix approach \cite{NijP:91,Nij:01}, group invariant solutions \cite{Mat:07,X:07}, and circle pattern interpretations \cite{AgaB:00}.

%It was first found \cite{NijP:91} that a discrete \Pa equation arises as a similarity reduction of the lattice KdV equation much in the same way continuous \Pa equations can be derived as self-similar reductions of integrable PDEs. Since then, uncovering the connection between the ABS equations and Sakai's classification of discrete \Pa equations became an area of intense interest in the integrable systems community. The main approaches for obtaining reductions have been periodic constraints \cite{GramaniP:91,Gramani:05,rcg:09}, similarity constraints derived by the infinite matrix approach \cite{NijP:91,Nij:01}, group invariant solutions \cite{X:07,Mat:07}, and circle pattern interpretations \cite{AgaB:00}.

In this paper we present a class of periodic reductions for the equations Q1, Q2, and Q3 in the ABS list \cite{abs:03}. As a sufficient condition for the resulting reductions to be at most second order in the independent variable, we consider here the reductions of the form $u(k,l+1)=m(u(k+1,l))$, where $m$ is a M\"obius transformation. The invertibility of the reduction constraint means that it can be expressed naturally on any staircase in the $\mathbb{Z}^2$ lattice. As we see in this paper, the condition that the resulting equation is second order allows us to use the algebro-geometric tools for mappings of the plane. The systematic approach allows us to give a complete description of reductions of the equations known as Q1, Q2, and Q3 (of the type $u(k,l+1)=m(u(k+1,l))$). Taking reductions that result in higher order mappings introduces several issues that the algebraic tools are not currently adequate to deal with. While some higher order reductions (a longer staircase periodicity condition) result in mappings that permit a reparametrisation into second order form, these are beyond the scope of this article. Similar M\"obius type reductions exist for Q4, the resulting equations are all non-linearisable, and calculating the deautonomisation proves too complex to be considered in this article.

The main result of this paper is a $q$-Painlev\'e equation resulting from a deautonomised reduction of Q3, which is presented in Section \ref{q3interesting}. The deautonomisation is provided by singularity confinement \cite{GramaniP:91}, which preserves the integrability of the reduction whilst allowing for the parameters to vary with the independent variable. The equation is shown to exist on the $A_1^{(1)}$ surface which belongs to the multiplicative type of surfaces in Sakai's classification of Painlev\'e systems \cite{sak:01}. To the authors' knowledge, this is the first example of such a equation being found as a reduction of an ABS equation.

For each reduction, we explore the integrability by examining the space of initial conditions, in the sense of \cite{oka:79,sak:01}. Initially we calculate the degree growth, or ``algebraic entropy", as introduced in \cite{bel:99}. When the degree growth is linear, indicating linearisability of the reduction, we use the geometry of the space of initial conditions to provide an explicit transformation of variables which linearises the system. 

In the remainder of Section \ref{introQreds}, we provide some definitions which will be used throughout the article. In Section \ref{admred}, we examine the set of reductions which are considered admissible for Q1, Q2 and Q3. In Section \ref{q1q2redsec}, we examine the reductions of Q1 and Q2, and show that in these cases all reductions are linearisable. We linearise the equations explicitly by constructing a birational transformation of the dependent variables which brings each reduction into a system of Riccati maps which may be solved in cascade, using the theory developed in \cite{tak:03}. This approach has the advantage of being an algorithmic process, in which the explicit linearisation can be found as opposed to ad hoc methods used previously, usually requiring guesswork based on knowledge of the linearisation of the autonomous forms of the equation, cf \cite{jos:06,rcg:09}. In Section \ref{q3redsec}, we will consider reductions from Q3, where we observe three reductions which are shown to be linearisable. In Section \ref{q3interesting}, we show that there exists a periodic reduction (an involution) of Q3 which is shown to be quadratic in degree growth. For this mapping we provide a deautonomisation via singularity confinement, and show that the configuration of the base points is that of the $A_1^{(1)}$ surface in the classification of Painlev\'e systems \cite{sak:01}. Finally, we make some concluding remarks in Section \ref{qredconc}.

\subsection{Space of initial conditions}\label{section:initialvaluespace}

For each reduction, we will investigate its integrability properties by constructing the space of initial conditions, as first introduced in \cite{oka:79} and developed as a classification for discrete Painlev\'e systems in \cite{sak:01}. When the number of base points is infinite, the mapping does not pass the singularity confinement test. However, for these mappings it is seen that the degree growth is linear and may be lifted to an ``analytically stable mapping", in the sense of \cite{tak:03}. Recast as a first order system in $\mathbb{P}^1\times\mathbb{P}^1$, we investigate the base points by blowing up. In the proceeding, we refer to the surface of initial conditions (base point-free blown up space) as $X$, and the mapping $(x,y)\rightarrow(\ol {x},\ol {y})$ as $\varphi$.

\begin{Def}
A base point is a point $(x,y)$ which is a singularity of the mapping $\varphi$ or its inverse, $\varphi^{-1}$. It may be identified as a point which is the blow down of, or blow up to a line, or an indeterminacy in the discrete vector field (the discrete phase space determined by the dynamical system).
\end{Def}

We use the notation where a blow up $\pi$ at the point $(x,y)=(a,b)\in\mathbb{C}^2$ is denoted by

\begin{equs}
\label{blowupdef}
(x,y)\stackrel{\pi}{\leftarrow} ((x-a)/(y-b),y-b)\cup (x-a,(y-b)/(x-a)).
\end{equs}
The exceptional divisor, $E$, is the inverse image of $(a,b)$ under the map $\pi$. We will often use the terms `exceptional divisor' and `blow up of the point' interchangeably.

Let $\pi: {\rm X} \to \mathbb{P}^1\times\mathbb{P}^1$ denote the blow up of $\mathbb{P}^1\times\mathbb{P}^1$ at $n$ base points $p_i$, $i=1,2,...,n$. Let the exceptional curve (curve with self-intersection -1 and birationally equivalent to $\mathbb{P}^1$) created by the blow up at $p_i$ be denoted by $e_i$ and let the linear equivalence classes of total transform of $x=$ constant and $y= $ constant be denoted by $h_x$ and $h_y$, respectively. From \cite{har:77} we know Pic($X$), the Picard group of $X$ is given by

\begin{displaymath}
{\rm Pic(\textit{X}\, )}=\mathbb{Z}h_x+\mathbb{Z}h_y+\sum_{i=1}^n\mathbb{Z}e_i.
\end{displaymath}

The intersection form $(|)$ is defined by 
\begin{equs}
 &(h_x|h_y)=(h_y|h_x)=1,\quad
 (h_x|h_x)=(h_y|h_y)=0,\\
 &(h_x|e_i)=(e_i|h_x)=(h_y|e_i)=(e_i|h_y)=0,\quad
 (e_i|e_j)=-\delta_{ij}.
\end{equs}
The anticanonical divisor of $X$ is given by
\begin{displaymath}
\delta=-\bm{K}_X=2h_x+2h_y-\sum_{i=1}^ne_i.
\end{displaymath}

%\begin{Def}
%For 2 dimensional systems of first order equations, such as the ones in this paper, we may express singularity confinement in the following way: Singularity confinement is the deautonomisation of the parameters which preserves the same linear mapping on the basis elements on the Picard group as in the case of the autonomous system.
%\end{Def}

\begin{Def} The algebraic entropy \cite{bel:99} of a mapping is given by the limit

\begin{equs}
\varepsilon = \lim_{n\rightarrow \infty} \frac{1}{n}\,{\rm log}\,d_n\,,
\end{equs}
where $d_n$ is the degree of the $n$-th iterate of the mapping in the initial data, in homogeneous coordinates.
\end{Def}

It is known \cite{via:06} that the algebraic entropy of the ABS equations is zero and the degree growth of their initial data is polynomial. For generic nonlinear equations, the algebraic entropy is greater than zero and the degree growth in their initial data of solutions is exponential. The decrease in the degree growth of the ABS equations comes from factorisations occurring in iterations of the equation. By performing a reduction, the family of solutions is restricted by the periodicity of the reduction, and thus we may find additional cancellations in the resulting mappings, possibly resulting in further decreased degree growth. Because of this, reductions from the integrable but not linearisable ABS equations may result in linearisable mappings.

%\begin{prop}
%Let $\ve_Q$ be the entropy of the ABS lattice equation $Q_i$, and $\ve_r$ be the entropy of the reduction. Then $\ve_r \leq \ve_Q$.
%\begin{proof}
%Assume $\ve_r > \ve_Q$, then there exists a periodic initial value problem for the lattice equation Q such that the entropy is $\ve_r$. By definition $\ve_Q$ is the maximal entropy for any IVP and thus $\ve_r \leq \ve_Q$, a contradiction.
%\end{proof}
%\end{prop}
%
%Thus all of the reductions obtained in this paper necessarily have zero entropy. 
We give a complete description of all the (integrable) reductions in the class we consider. It was observed \cite{ram:00} that mappings having low growth (either no growth or linear growth) can be linearised. Hence for each map a calculation of the degrees of the first few iterates is a good test to determine the linearisability of a mapping, we compute these for the reductions obtained in this paper.

\section{Admissible reductions}\label{admred}

This paper will consider three Q type quad-equations, namely Q1, Q2, Q3,
in the ABS classification \cite{abs:03} of integrable lattice
equations possessing 3D consistency. These quad equations take the form
$Q(u,\ol {u},\hat{u},\hat{\ol {u}})=0,$
where $Q(u,\ol {u},\hat{u},\hat{\ol {u}})$ is an irreducible multi-affine polynomial in
$u,\ol {u},\hat{u},\hat{\ol {u}}$.

\begin{equs}
\label{}
   Q3_\delta &= {\rm sinh}(\alpha)(u\ol {u}+\hat{u}\hat{\ol {u}})-{\rm sinh}(\beta)(u\hat{u}+\ol {u}\hat{\ol {u}}) \\
   &\hspace{1in}- {\rm sinh}(\alpha-\beta)(\hat{u}\ol {u}+u\hat{\ol {u}}-\delta^2{\rm sinh}(\alpha){\rm sinh}(\beta))\,,\\
   Q2_\delta &=\alpha(u-\hat{u})(\ol {u}-\hat{\ol {u}})-\beta(u-\ol {u})(\hat{u}-\hat{\ol {u}})+\alpha\beta(\alpha-\beta)(u+\hat{u}+\ol {u}+\hat{\ol {u}}-\alpha^2-\beta^2+\alpha\beta)\,,\\
   Q1_\delta &= \alpha(u-\hat{u})(\ol {u}-\hat{\ol {u}})-\beta(u-\ol {u})(\hat{u}-\hat{\ol {u}})+\delta^2\alpha\beta(\alpha-\beta)\,,
\end{equs} 
where the parametrisation of $Q3_\delta$ is due to Hietarinta \cite{hie:05}. Here $u=u(k,l)$, and the hat/bar notation used throughout is $\ol {u}=u(k+1,l)$, $\hat{u}=u(k,l+1)$.

We investigate all possible reductions of the form $\hat{u}=m(\ol {u})$, where $m(u)=(a_1 u+a_2)/(a_3 u+a_4)$, 
is a M\"obius transformation, where
$a_1$, ..., $a_4$ are independent of $k$ and $l$. We call a reduction \textit{admissible} if the reduced equation is consistent on the lattice, i.e. the solution to the reduction is M\"obius invariant on the lattice. More specifically where the ABS equation in question is written $Q(u, \ol {u}, \hat{u}, \hat{\ol {u}})=0$, the reduced equation on the quadrilateral $Q=Q(u,\ol {u},\hat{u},\hat{\ol {u}})=Q(u,\ol {u},m(\ol {u}),m(\ol {\ol {u}}))$ must be equal (up to a nonzero multiplicative factor) to the reduced equation on the quadrilateral $\underline{\widehat{Q}}=Q(\underline{\hat{u}},\hat{u},\underline{\hat{\hat{u}}},\hat{\hat{u}})=Q(m(u),m(\ol {u}),m(m(\ol {u})),m(m(\ol {\ol {u}})))$. In this way, the reduced equation (without loss of generality written in terms of bars rather than hats) is the same on any infinite (1,1) staircase on the lattice. A (1,1) staircase is a connected path which is non-decreasing, or non-increasing on the $\mathbb{Z}^2$ lattice along which the initial conditions and periodic condition $\hat{u}=m(\ol {u})$ are satisfied \cite{kq10}. While, in the past, the (1,1) reduction is used to describe the periodic condition $\hat{u}=\ol {u}$, here it is more appropriate to use the term for the reductions of the type $\hat{u}=m(\ol {u})$. In this paper, $Q$ is written as shorthand for the quadrilateral equation which depends on the four vertices of the quadrilateral.

\subsection{Reductions}
By following the above procedure, we find that the admissible reductions $\hat{u}=m(\ol {u})$ for the equation $Q1_\delta$ are

\begin{equs}
m(u)  &=a\pm u\,, \text{for any } \delta \,,  \\
m(u)  &=\frac{a_1 u+a_2}{a_3 u+a_4}, {\rm when}\,\, \delta=0\,,   
\end{equs}
provided $a_2a_3-a_1 a_4\neq0$.

The only admissible reduction $\hat{u}=m(\ol {u})$ for the equation $Q2$ is

\begin{equs}
\label{}
  m(u)  &=u\,.
\end{equs}

The admissible reductions $\hat{u}=m(\ol {u})$ for the equation $Q3_\delta$ are

\begin{equs}
\label{}
  m(u)  &=\pm u\,, \text{for any } \delta \,,  \\
  m(u)  &=\frac{a}{u}\, {\rm when}\,\, \delta=0\,,\\
    m(u)  &=c\, u\, {\rm when}\,\, \delta=0\,. 
\end{equs}

The reductions $\hat{u}=\ol {u}$ of Q2 and Q3 were carried out in \cite{jos:06,rcg:09}. The resulting mappings were shown to be linearisable. It is expected that higher order reductions could lead to mappings that are not linearisable, whose nonautonomous forms will be related to discrete Painlev\'e equations. As previously mentioned, the issue with taking higher order reductions is identifying the reduced equation. In some cases \cite{Hay:07,Orm:12}, one can integrate the reduction to a second order difference equation, but this proves difficult in general. The process of identifying a second order equation is problematic as one requires either that the equation is in a known form, or a transformation can be found that brings it into a known form, which is practically difficult. If one attempts to identify an equation by taking an appropriate continuous limit, one is faced with the problem that many discrete equations have the same continuous limit. Two equations can not be transformed to each other unless they share the same surface of initial conditions and their actions on the Picard group are conjugate to each other \cite{Sakai:07}. Only then is it theoretically possible to find a change of variables which will bring one equation to the other. For this reason, we study the surface of initial conditions and its action on the Picard group of that surface, so as to identify the equation in a way that is independent of its presented form.

For those cases where $m$ is an involution, in taking reductions $\hat{u}=m(\ol {u})$, we are effectively taking a sub-case of a higher order (2,2) reduction. The bonus is that instead of resulting in a fourth order equation which would need to be integrated twice (if possible) to bring it into a identifiable form, the equation is automatically second order. This allows one to use the algebro-geometric tools to identify the geometry of the reductions.

\section{Reductions of Q1 and Q2}\label{q1q2redsec}

In this section, we examine the admissible reductions for the Q1 and Q2 equations, as stated in Section \ref{admred}. In each case we show the reduced equation, some of which are clearly linear or linearisable, while others are not. In those cases where linearisation is not immediately obvious, the exact degree growth is calculated through the realisation of the space of initial conditions \cite{tak:01}. Although infinitely many blow ups are required for non-confined singularities, we may still obtain a great wealth of information from their explicit construction. That is, we can rigorously determine growth properties, as well as find pencils of invariant curves which determine linearising transformations by the theory developed in \cite{tak:03}.

\subsection{Q1: Where $\hat{u}=a+\ol {u}$}

This leads to the equation

\begin{equs}
\label{xhisxb}
& \alpha(u-\ol {u}-a)(\ol {u}-\ol {\ol {u}}-a)+\beta(u-\ol {u})(\ol {u}-\ol {\ol {u}})+\delta^2\alpha\beta(\alpha-\beta)=0.
\end{equs}
Upon the substitution $y=u-\ol {u}$ we have

\begin{equs}
\label{}
\alpha(y-a)(\ol {y}-a)+\beta y\ol {y} +\delta^2\alpha\beta(\alpha-\beta)=0,
\end{equs}
and as this is a discrete Riccati equation, the general solution for $y$ can be easily obtained.

\subsection{Q1: Where $\hat{u}=a-\ol {u}$}\label{q1xbisxhsec}

This leads to the equation

\begin{equs}
\label{}
\alpha(u+\ol {u}-a)(\ol {u}+\ol {\ol {u}}-a)+\beta(u-\ol {u})(\ol {u}-\ol {\ol {u}})+\delta^2\alpha\beta(\alpha-\beta)=0,
\end{equs}
where the $a$ can be removed by a shift $u\mapsto u+a/2$, resulting in

\begin{equs}
\label{xhismxb}
\alpha(u+\ol {u})(\ol {u}+\ol {\ol {u}})+\beta(u-\ol {u})(\ol {u}-\ol {\ol {u}})+\delta^2\alpha\beta(\alpha-\beta)=0.
\end{equs}

By calculating the first degrees of (\ref{xhismxb}) in the iterates of the mapping, we obtain the degree growth sequence  2, 4, 6, 8, 10, ..., which would suggest the mapping has linear degree growth, and thus is linearisable. It is not immediately clear what transformation will linearise (\ref{xhismxb}) as was the case in (\ref{xhisxb}). To find the linearising transformation, and to confirm the trend in degree growth, we construct the space of initial conditions, in the sense of \cite{sak:01}. Equation (\ref{xhismxb}) is equivalent to the system

\begin{equs}
\label{xhismxbsys}
  \ol {x}  &= \frac{\delta^2\alpha\beta(\beta-\alpha)-\alpha x (x+y)+\beta x (x-y)}{\alpha  (x+y)+\beta (x-y)}\, , \\
  \ol {y}  & = x\,,
\end{equs}
where $x = \ol {u}$ and $y = u$. In $\mathbb{P}^1\times\mathbb{P}^1$, this system has base points at

\begin{equs}
				\label{}
				p_1:&\, (x,y)  =\left(\frac{\delta(\alpha-\beta)}{2} ,\frac{-\delta(\alpha+\beta)}{2}  \right) \, , 
\,\,\,p_2: \,(x,y)  =\left(\frac{-\delta(\alpha-\beta)}{2} ,\frac{\delta(\alpha+\beta)}{2}  \right) \,, \\
p_3:&\,(x,y)  =\left(\frac{-\delta(\alpha+\beta)}{2} ,\frac{\delta(\alpha-\beta)}{2}  \right) \,, \,\,\,
p_4:\,(x,y)  =\left(\frac{\delta(\alpha+\beta)}{2} ,\frac{-\delta(\alpha-\beta)}{2}  \right) \,, \\
p_5:&\,(1/x,1/y)  =\left( 0, 0  \right) \,.
\end{equs}

The surface $Y$ obtained by the blow up, $\pi$, of $\mathbb{P}^1\times\mathbb{P}^1$ at $p_i$, $i=1,...,5$, is free of indeterminacy as a mapping from $Y$ to $\mathbb{P}^1\times\mathbb{P}^1$. We call $F$ the blow up of $p_1$, and similarly we call $G,f,g,E$ the blow ups of $p_2, p_3,p_4,p_5$ respectively, i.e.

\begin{equs}
\label{bpsq1red}
p_1 & \stackrel{\pi}{\leftarrow} F\,, \,\,
p_2  \stackrel{\pi}{\leftarrow} G\,, \,\,
p_3  \stackrel{\pi}{\leftarrow} f\,, \,\,
p_4  \stackrel{\pi}{\leftarrow} g\,, \,\,
p_5  \stackrel{\pi}{\leftarrow} E\,.
\end{equs}

We find however that the lines $h_y-E$ (given by the relation $1/y=0$) and $h_x-E$ (given by the relation $1/x=0$) are blown down by the mapping in the following way:

\begin{equs}
\label{first_calc_eqns}
h_x-E\stackrel{\varphi}{\rightarrow} &\,(y/x,1/y)  =\left(\frac{\alpha+\beta}{\beta-\alpha},0\right)\,:=p_6,\\
h_y-E\stackrel{\varphi^{-1}}{\rightarrow} &\,(1/x,x/y)  =\left(0,\frac{\alpha+\beta}{\beta-\alpha}\right)\,:=p_7.
\end{equs}

We demonstrate the procedure for the first of (\ref{first_calc_eqns}) as follows: The line $h_x - E$ is parametrized by the equations $(X = 1/x = 0, y = y)$, upon substitution into (\ref{xhismxbsys}) we have $(\ol {x}, \ol {y}) = (\infty, \infty) = p_5$. The image of $p_5$ in $Y$ is the exceptional line $E$, coordinatized by $(y/x, 1/y) = (u, 0)$. The image of $h_x - E$ on $E$ is given by $(\ol {y}/\ol {x}, 0)$ where $(X = 1/x = 0, y = y)$. Here

\begin{equs}
\ol {y}/\ol {x} &= -\frac{\alpha(1+Xy) + \beta(1-Xy)} {\alpha (1+Xy-\beta^2\delta^2 X^2) - \beta(1 - Xy - \alpha^2\delta^2X^2)} |_{X=0}\, ,\\
& = \frac{\alpha+\beta}{\beta-\alpha}\,,
\end{equs}
and hence we arrive at the first of (\ref{first_calc_eqns}).

Upon the blow up of $p_6\stackrel{\pi}{\leftarrow} b_1$ and $p_7\stackrel{\pi}{\leftarrow} d_1$, we find that these exceptional lines $b_1$ and $d_1$ are blown down in the following way:

\begin{equs}
\label{}
b_1\stackrel{\varphi}{\rightarrow} &\,(y/x,1/y)  =\left(0,0\right)\,:=p_8,\\
d_1\stackrel{\varphi^{-1}}{\rightarrow} &\,(1/x,x/y)  =\left(0,0\right)\,:=p_9.
\end{equs}
The first of these can been seen as the line $b_1$ is parametrized by the relations 
\begin{equs}
(u, v) = \left(\frac{y((\alpha+\beta)x+(\alpha-\beta)y)}{(\alpha-\beta)x}, \frac{1}{y}\right) = (u, 0)\,,
\end{equs}
and we find the image of $v$=$0$, $u$=$u$ in $Y$ is the point on the exceptional line $E$ given by $(y/x,1/y)  =\left(0,0\right)$.

\begin{figure}
\begin{center}
\includegraphics[width=4in]{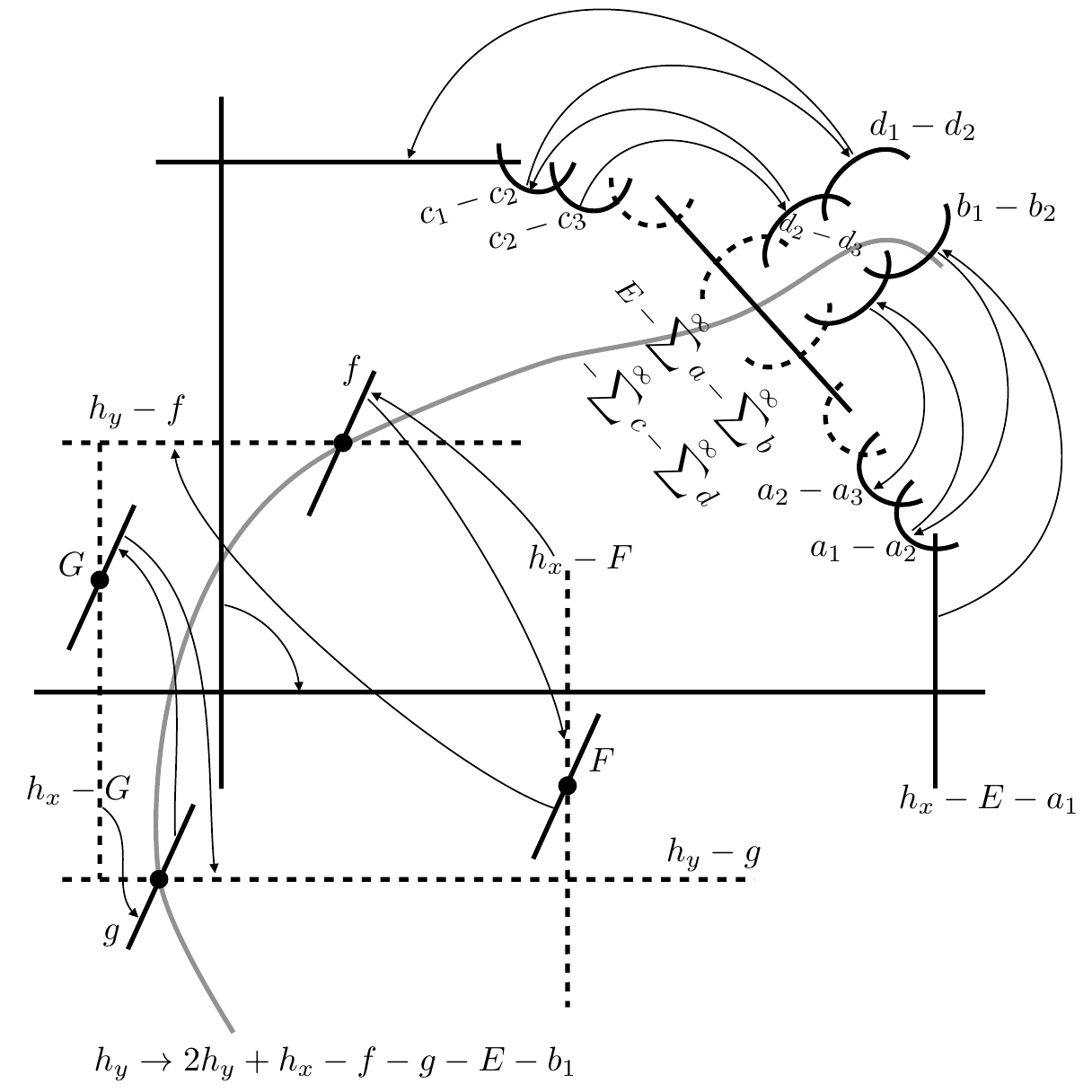}
\caption{The action of the system (\ref{xhismxbsys}) on the elements of the Picard group. Dotted curved arcs represent infinitely many of the same type of blow up.}
\label{xbismxhfig1}
\end{center}
\end{figure}

These points are the blow downs of exceptional lines, and so we must blow up $p_8\stackrel{\pi}{\leftarrow} a_1$ and $p_9\stackrel{\pi}{\leftarrow} c_1$. In turn, the iterations of these lines are blown down to points, and we find we require an infinite sequence of blow ups to regularise the mapping, this is a case of unconfined singularities. Nevertheless, we find that the sequence is easily described, as in Figure \ref{xbismxhfig1}.

The action of the mapping on elements of the Picard group is shown in Figure \ref{xbismxhfig1}. The Picard group is given by 

\begin{displaymath}
{\rm Pic(X)}=\mathbb{Z}h_x+\mathbb{Z}h_y+\mathbb{Z}E+\mathbb{Z}f+\mathbb{Z}F+\mathbb{Z}g+\mathbb{Z}G+\sum_{i=1}^\infty \mathbb{Z}a_i+\sum_{i=1}^\infty \mathbb{Z}b_i+\sum_{i=1}^\infty \mathbb{Z}c_i+\sum_{i=1}^\infty \mathbb{Z}d_i\,,
\end{displaymath}

The actions on the generators of the Picard group are as follows

\begin{equs}
\label{q1redpicmap}
 & h_y \mapsto \,  2h_y+h_x-E-f-g-b_1\,, \,\,
  h_x \mapsto \,  h_y \,, \\
 & E \mapsto \,  h_y-b_1\,, \,\,
  f \mapsto \,  F \,, \,\,
  g \mapsto \,  G\,, \,\,
  F \mapsto \,  h_y-f\,, \,\,
  G \mapsto \,  h_y-g\,, \\
  &  d_i \mapsto \,  c_{i-1} \,, \, i\geq2\,,\\
 & b_i \mapsto \,  a_i \,,\,\,
  a_i \mapsto \, b_{i+1} \,, \,\,
  c_i \mapsto \,  d_i \,, \, i\geq1\,, \\
 & d_1 \mapsto \,  h_y-E \,.  
\end{equs}
We can calculate the above directly by substituting in the lines and finding the resulting image. For example, the generic line $y$= constant = $c$, denoted $h_{y=c}$,  has the image

\begin{equs}
  \ol {x}  &= \frac{\delta^2\alpha\beta(\beta-\alpha)-\alpha x (x+c)+\beta x (x-c)}{\alpha  (x+c)+\beta (x-c)}\, , \\
  \ol {y}  & = x\,,
\end{equs}
for which $x$ can be simply eliminated. Direct substitution (in the relevant coordinate charts) shows that the image curve intersects the lines $E, f, g, b_1$ exactly once each, and also it intersects the lines $h_y$ twice and $h_x$ once (being quadratic in $\ol {y}$ and monic in $\ol {x}$).

The mapping (\ref{q1redpicmap}) is a linear system, and so we can diagonalise to find

\begin{equs}
\label{}
 &\varphi^n(h_y)\\\nonumber
  &=2n h_y+2(n-1)h_x-n(f+g)-(n-1)(F+G)-2(n-1)E-2b_1 + (a_i,\,b_i \,{\rm terms}). 
\end{equs}

The degree of the $n$-th iterate in the initial data, ($x_0,y_0$), is given by the maximum of the coefficients of $h_y$ or $h_x$ in $\varphi^n(h_y)$ or $\varphi^n(h_x)$. These coefficients are equal to the intersection of a generic degree one initial curve with the lines $x=$ constant or $y=$ constant, which is equal to the degree of the mapping, by B\'ezout's theorem. Given that $\varphi^n(h_x)=\varphi^{n-1}(h_y)$, and the degree is increasing, the maximum degree is given by the largest of the coefficients of $h_x$ and $h_y$ in $\varphi^n(h_y)$. In this case the degree of the $n$-th iterate of the mapping is indeed $2n$ (the coefficient of $h_y$), which agrees with the numerical evidence. 

Although the system (\ref{xhismxbsys}) has infinitely many base points, we can lift the mapping to an ``analytically stable system" in the sense of \cite{tak:03}. An analytically stable system is one in which there are no singularity patterns of the type $e\mapsto\,{\rm sequence\, of\, points}\mapsto e$ where $e$ is an effective divisor. In our case we find this is possible by only blowing up the points resulting in the curves $f,\,F,\,g,\,G,\,E$, and then

\begin{equs}
\label{}
  h_y \mapsto \, & 2h_y+h_x-E-f-g\,,  \\
  h_x \mapsto \, & h_y \,, \\
  E \mapsto \, & h_y\,, \\
  f \mapsto \, & F \,, \\
  g \mapsto \, & G\,, \\
  F \mapsto \, & h_y-f\,, \\
  G \mapsto \, & h_y-g\,.
\end{equs}
This action is a linear system, with a non-trivial eigenspace with the eigenvalue 1. It leaves invariant a family of curves, C, whose representation in the Picard group is 

\begin{equs}
\label{invcurveq1}
C=2h_y+2h_x-f-g-F-G-2E\,,
\end{equs}
that is $\varphi(C)=C$. A generic degree 2 curve in $x$ and $y$ is given by 
\begin{equs}
\label{}
c_{22}x^2y^2 + c_{12} xy^2 + c_{21}x^2y +c_{20}x^2 + c_{02}y^2 + c_{11} xy +c_{10}x+c_{01}y+ c_{00} = 0\,,
\end{equs}
however since $C=2h_y+2h_x-f-g-F-G-2E$, $C$ is a curve of degree 2 in both $x$ and $y$, which passes through the points (\ref{bpsq1red}) $p_1,...,p_4$ and $p_5$ with multiplicity 1, 1, 1, 1 and 2 respectively. We use this fact to reduce the number of coefficients $c_{ij}$. This family is described by the pencil of curves

\begin{equs}
\label{}
c_1(\alpha^2(x+y)^2-\beta^2(x-y)^2)-c_2(\delta^2(\alpha^2+\beta^2)-2(x^2+y^2))=0\,.
\end{equs}
The ratio $c_1/c_2$ defines a pencil of curves isomorphic to $\mathbb{P}^1$. If we define

\begin{equs}
\label{q1xhismblin}
   v &:=\frac{c_1}{c_2} = \frac{\delta^2(\alpha^2+\beta^2)-2(x^2+y^2)}{\alpha^2(x+y)^2-\beta^2(x-y)^2}\,,   \\
   w &:=\frac{y+\frac{\delta(\alpha+\beta)}{2}}{x-\frac{\delta(\alpha-\beta)}{2}}  \,,
\end{equs}
where we have chosen $w$ such that $w$=constant intersects $v$=constant precisely once, so that the inverse transform is unique. The transformation, $w$, is found by finding another family of curves in $\PX$, isomorphic to $\mathbb{P}^1$ (and thus has self intersection 0), which gives a single intersection with (\ref{invcurveq1}). One candidate is $W=h_x+h_y-f-E$, which results in $w$. Hence the linearisation (\ref{q1xhismblin}) effectively maps the coordinates $(x,y)\in \pxp$ to $(v,w) \in S \simeq \pxp$, where the dynamics on $v$ are simple, given by $h_v \mapsto h_v$. By taking the upshift of (\ref{q1xhismblin}), substituting in (\ref{xhismxbsys}) and then substituting in the inverse transform for $x$ and $y$ in terms of $v$ and $w$, the system (\ref{xhismxbsys}) is written as a cascade of discrete Riccati equations which can be linearised in the standard way.

\begin{equs}
\label{linearq1red}
  \ol {v}  &= \frac{\al^2-\be^2+2\al^2\be^2 v}{\al^2\be^2((\al^2-\be^2)v+2)}\,,  \\
  \ol {w}  &= \frac{f_1 vw+f_{2,+}v+f_3w+f_4}{-f_1 vw-f_{2,-}v-f_3w+f_4}\,,  
\end{equs}
where

\begin{equs}
\label{}
  f_1  &=\al(\al^2-\be^2)\,,   \\
  f_{2,\pm}  &=\al(\al\pm\be)^2\,,   \\
  f_3  &=2\al\,,   \\
  f_4  &=2\be\,.
\end{equs}
Equations (\ref{linearq1red}) are two discrete Riccati type equations which may be solved in cascade.

\subsection{Q1: Where $\hat{u}=\frac{a_1 \ol {u}+a_2}{a_3 \ol {u}+a_4}\,,\, \delta=0$, $a_2a_3-a_1 a_4\neq0$} $\,$ \label{q1lastsect}

This leads to a cumbersome equation which has degree growth 1,2,3,4,5,6,..., i.e. linear growth. To spare the reader unnecessarily complex calculations, in this Section we will consider only the sub-case $\hat{u}=a/\ol {u}$, which retains the essential features of the full reduction. This far simpler looking equation reads

\begin{equs}
\label{q1xhisaonxb}
\alpha(u\ol {u}-a)(\ol {u}\ol {\ol {u}}-a)-a\beta(u-\ol {u})(\ol {\ol {u}}-\ol {u})=0\,.
\end{equs}
Again the degree growth is 1,2,3,4,5,6,..., which suggests linear growth. As before we verify this by calculating the degree growth exactly and finding the linearising transform through the blown up space of initial conditions.

Let $x = \ol {u}$ and $y = u$ so we recast (\ref{q1xhisaonxb}) as the system $\varphi :(x,y)\mapsto (\ol {x},\ol {y})$, where $(\ol {x},\ol {y})$ are given by

\begin{equs}
\label{q1xhisaonxbsys}
\ol {x}=&\frac{c^2((\beta-\alpha)xy+\alpha c^2-\beta x^2)}{c^2(\alpha-\beta)x+\beta c^2y-\alpha x^2y}\,,\\
\ol {y}=&x\,,
\end{equs}
where $c^2=a$ in (\ref{q1xhisaonxb}).

\begin{figure}
\begin{center}
\includegraphics[width=4in]{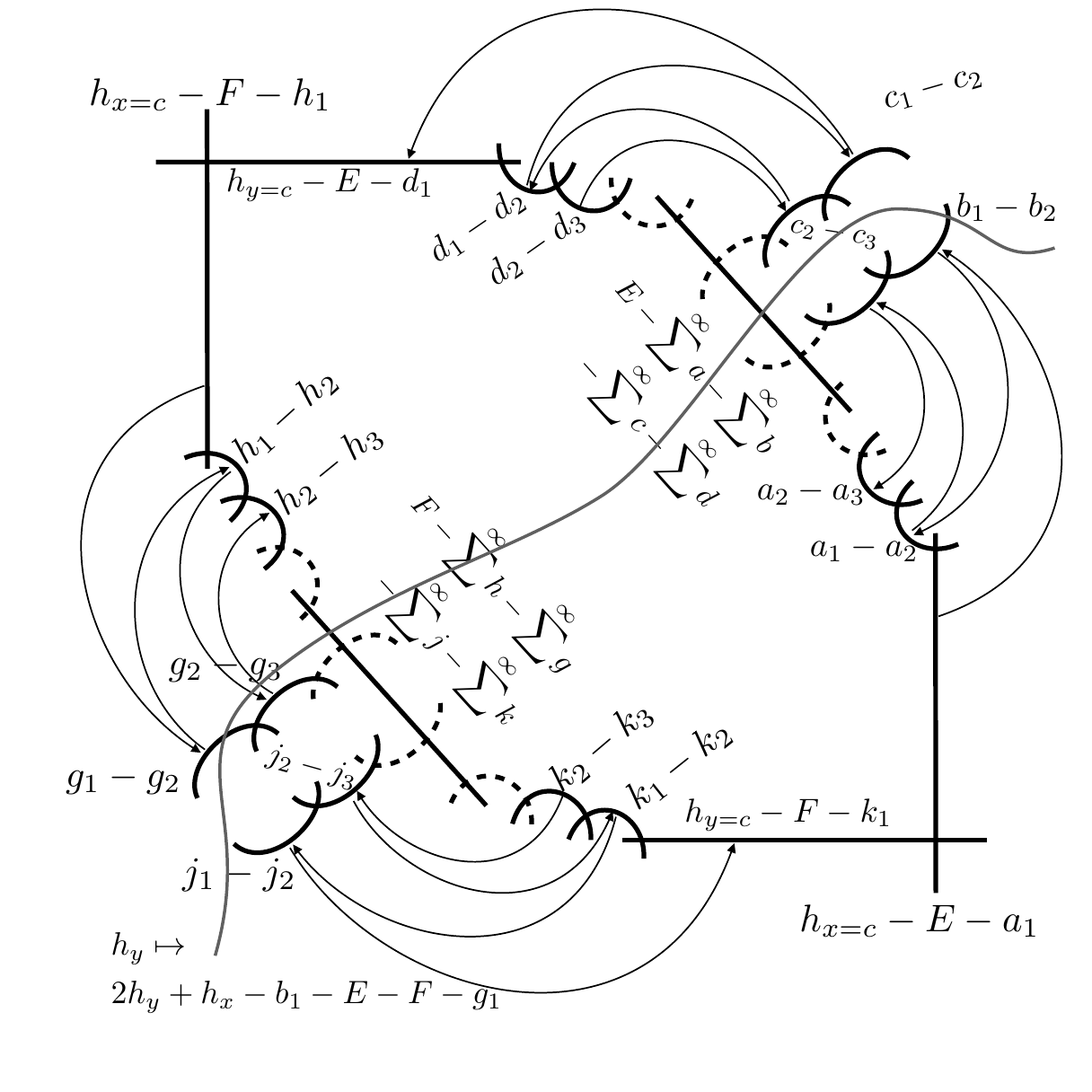}
\caption{The action of the system (\ref{q1xhisaonxbsys}) on the elements of the Picard group. Dotted curved arcs represent infinitely many of the same type of blow up.}
\label{q1xhisaonxbfig1}
\end{center}
\end{figure}

This mapping is indeterminate at the points $(x,y)=(\pm c,\pm c)$. The blow up $\pi$ at $(c,c)\stackrel{\pi}{\leftarrow} E$ and $(-c,-c) \stackrel{\pi}{\leftarrow} F$ lifts the mapping $\varphi$ to a analytically stable mapping. There are, however, further indeterminate points, which arise due to the blow downs of the lines $x=\pm c$ by $\varphi$ and $y=\pm c$ by $\varphi^{-1}$, which in turn result in an infinite sequence of blow downs as in Section \ref{q1xbisxhsec}. The explicit sequence can be observed in Figure \ref{q1xhisaonxbfig1}.

The action of the mapping on elements of the Picard group are shown in Figure \ref{q1xhisaonxbfig1}. The actions on the generators of the Picard group are as follows

\begin{equs}
\label{q1red2pic}
&  h_y \mapsto \,  2h_y+h_x-E-b_1-F-g_1\,, \,\,
  h_x \mapsto \,  h_y \,, 
  E \mapsto \,  h_y-b_1\,, \,\,\\
 & a_i \mapsto \, b_{i+1} \,,\,\,
  b_i \mapsto \,  a_i \,, \, \,\,
  d_i \mapsto \,  c_i \,, \, i\geq1\,,\\
 & c_1 \mapsto \,  h_y-E \,, \,\,
  F \mapsto \,  h_y-g_1 \,, \,\,
  j_1 \mapsto \,  h_y-F \,, \\
&  g_i \mapsto \, h_i \,,\,\,\,
  h_i \mapsto \,  g_{i+1} \,, \,\,
  k_i \mapsto \,  j_i \,, \, i\geq1\,,\\
 & c_i \mapsto \,  d_{i-1} \,, \,\,\,
  j_i \mapsto \,  k_{i-1} \,, \,i\geq2\,.
\end{equs}
From (\ref{q1red2pic}) we find

\begin{equs}
\label{}
 \varphi^n(h_y) &=(n+1) h_y+nh_x-n(E+F)-(a,b,g,h\, {\rm terms}).
\end{equs}
Thus the degree of the $n$-th iterate of the mapping is indeed $n+1$ (the coefficient of $h_y$). 

The analytically stable map obtained by blowing up $\mathbb{P}^1\times\mathbb{P}^1$ at the points resulting in $E$ and $F$ will give a linearising transformation. In this case we find 

\begin{equs}
\label{asq1xhisaonxb}
  h_y \mapsto \, & 2h_y+h_x-E-F\,,  \\
  h_x \mapsto \, & h_y \,, \\
  e \mapsto \, & h_y\,, \\
  f \mapsto \, & h_y \,.
\end{equs}
From (\ref{asq1xhisaonxb}), there exists the invariant class of divisors 

\begin{equs}
\label{}
    &  h_y+h_x-E-F\,,
\end{equs}
which is representative of the pencil of curves (isomorphic to $\mathbb{P}^1$)

\begin{equs}
\label{}
c_1 c^2 (y-x)+c_2(c^2-xy)=0\,.
\end{equs}
By choosing 

\begin{equs}
\label{lineariseq1xhisaonxb}
  s = &\frac{c_2}{c_1}=\frac{c^2(x-y)}{c^2-xy}\,,   \\
  t = &  x\,,
\end{equs}
the resulting equations for $s$ and $t$ are

\begin{equs}
\label{}
  \ol {s}  &=-\frac{\alpha c^2}{\beta s}\,,   \\
 \ol {t}   & =\frac{\alpha c^2-\beta s t}{\alpha t-\beta s}\,. 
\end{equs}
The equation for $s$ is trivially solved and the equation for $t$ is just a discrete Riccati equation. Thus the system (\ref{q1xhisaonxbsys}) is linearised by (\ref{lineariseq1xhisaonxb}).

\subsection{Q2, where $\hat{u}=\ol {u}$}
The reduction $\hat{u}=\ol {u}$ was covered in \cite{jos:06}, where it was shown to be linearisable. 

\section{Reductions from Q3}\label{q3redsec}

\subsection{Q3, where $\hat{u}=\ol {u}$}
This was discussed in \cite{jos:06}, where it was shown to be linearisable. The nonautonomous form of this reduction, deduced by singularity confinement, was found in \cite{rcg:09}, where it was also shown to be linearisable.

\subsection{Q3, where $\hat{u}=-\ol {u}$}

This results in the equation

\begin{equs}
\label{q3xhismxb}
({\rm sinh}\,\alpha+{\rm sinh}\,\beta)(u\ol {u}+\ol {u}\ol {\ol {u}})+{\rm sinh}\,(\alpha-\beta)(u\ol {\ol {u}}+\ol {u}^2+\delta^2{\rm sinh}\,\alpha\,{\rm sinh}\,\beta)=0\,.
\end{equs}
Equation (\ref{q3xhismxb}) for $\delta\neq 0$, has degree growth 1, 2, 4, 6, 8, 10, 12, 14, ..., whereas for $\delta=0$ the degree growth is 1, 2, 3, 4, 5, 6, 7, ..., which suggests linearisability (owing to linear growth) in both cases. In $\mathbb{P}^1\times\mathbb{P}^1$, on letting $x=\ol u$, $y=u$ we recast the equation (\ref{q3xhismxb}) as the system $\varphi : (x,y)\mapsto (\ol {x},\ol {y})$, where

\begin{equs}
\label{q3xhismxbsys1}
 \ol {x}   &= -\frac{xy({\rm sinh}(\ga-z)+{\rm sinh}(\ga+z))-{\rm sinh}(2z)x^2+\delta^2 {\rm sinh}(\ga-z){\rm sinh}(\ga+z)}{x({\rm sinh}(\ga-z)+{\rm sinh}(\ga+z))-y{\rm sinh}(2z)}\,,   \\
   \ol {y} &  =x\,,
\end{equs}
where we have made the change of variables $\al\rightarrow \ga-z$, $\be \rightarrow \ga+z$. In $\mathbb{P}^1\times\mathbb{P}^1$, this system has base points at

\begin{equs}
\label{}
p_1:&\, (x,y)  =({\rm sinh}\,z,{\rm sinh}\,\ga) \, , \\
p_2:& \,(x,y)  =(-{\rm sinh}\,z,-{\rm sinh}\,\ga) \, , \\
p_3:&\,(x,y)  =({\rm sinh}\,\ga,{\rm sinh}\,z) \, , \\
p_4:&\,(x,y)  =(-{\rm sinh}\,\ga,-{\rm sinh}\,z) \, , \\
p_5:&\,(1/x,1/y)  =\left( 0, 0  \right) \,.
\end{equs}

Interestingly, from a geometric standpoint, this set of base points is the same as that of the mapping of Section \ref{q1xbisxhsec}. In fact, geometrically we have the same infinite sequence of blow ups required to regularise the space of initial conditions. The action on the Picard group is the same as that of Section \ref{q1xbisxhsec}, and so we omit the calculations. We find that the pencil of curves

\begin{equs}
\label{}
c_1\delta^2({\rm sinh}\,\gamma\, x-{\rm sinh}\,z\,y)({\rm sinh}\,\gamma\, y-{\rm sinh}\,z\,x)+c_2\left[({\rm sinh}^2\,\gamma +{\rm sinh}^2\,z)\delta^2-(x^2+y^2)   \right]=0\,
\end{equs}
is preserved by the system (\ref{q3xhismxbsys1}). If we put 

\begin{equs}
\label{lineariseq3xhismxb}
   s &=\frac{c_1}{c_2}=\frac{x^2+y^2-\delta^2\left(\sinh ^2\gamma+\sinh ^2z\right)}{\delta^2 (x \sinh z-y \sinh \gamma) (y \sinh z-x \sinh \gamma)} \,,  \\
   t & =\frac{y-\delta \, {\rm sinh}\,\gamma}{x-\delta \,{\rm sinh}\,z}\,,
\end{equs}
then
\begin{equs}
\label{q3mlinear}
   \ol {s} &=\frac{16 \sinh (z) \sinh (\gamma )-s \delta ^2 (\cosh (2 \gamma )-\cosh (2 z))^2}{\delta ^2 (\cosh (2 \gamma )-\cosh (2 z))^2 \left(s \delta ^2 \sinh (z) \sinh
   (\gamma )+1\right)} \,,  \\
    \ol {t}&=\frac{a_0+a_1t}{a_2+a_3t}\,,
\end{equs}
where

\begin{equs}
\label{}
a_0 &=  \sinh (z) \left(s \delta ^2 \cosh (2 \gamma )-s \delta ^2+2\right)-2 s \delta ^2 \sinh ^3(\gamma )+2 \sinh (\gamma )\,,  \\
a_1& = (\sinh (\gamma )-\sinh (z)) \left(2 s \delta ^2 \sinh (z) \sinh (\gamma )+2\right)\,,\\
a_2 &=\sinh (z) \left(-s \delta ^2 (\cosh (2 z)-2 \sinh (z) \sinh (\gamma ))+s \delta ^2+2\right)+2 \sinh (\gamma )\,,\\
a_3 &=2 (\sinh (z)-\sinh (\gamma )) \left(s \delta ^2 \sinh (z) \sinh (\gamma )+1\right)\,.
\end{equs}
The system (\ref{q3mlinear}) can be solved in cascade, as Riccati equations. The case where $\delta = 0$ causes (\ref{lineariseq3xhismxb}) to not be well defined, however this is a sub case ($c=-1$) in the following section.

\subsection{Q3, where $\hat{u}=c \ol {u}$, $\delta=0$}

This results in the equation

\begin{equs}
\label{q3xiscx}
 \sinh (\alpha ) {u} \left(c^2 {\xa}+\xub\right)-c  \sinh (\beta ) {u}(\xub+\xa)-c \sinh (\alpha -\beta ) \left(\xub
   {\xa}+{u}^2\right)=0\,.
\end{equs}

This equation has degree growth 1,2,3,4,5,..., and as a system where $y=\xub$, has a similar singularity structure to the equation in Section \ref{q1lastsect}, where the base points here are at zero and infinity, rather that $\pm c$. Hence a similar calculation of the linearisation ensues, and the linearising transform can be calculated in this way. However, by observation, if we divide (\ref{q3xiscx}) by $u\xub$ and let $w=u/\xub$, then we arrive at

\begin{equs}
\label{q3xiscxint}
 \sinh (\alpha ) \left(c^2 \ol {w}w+1\right)-c  \sinh (\beta ) ( \ol {w}w+1)-c \sinh (\alpha -\beta ) \left( \ol {w}
   +{w}\right)=0\,,
\end{equs}
which is a discrete Riccati equation.

%%%%%%%%%%%%%%%%%%%%%%%%%%%%%%%%%%%%%%%%%%%%%%%%%%%%%%%%%%%
%% 5. Q3, where $\hat{u}=a/\ol {u}$
%%%%%%%%%%%%%%%%%%%%%%%%%%%%%%%%%%%%%%%%%%%%%%%%%%%%%%%%%%%
\section{Q3, where $\hat{u}=a/\ol {u}$, $\delta=0$}\label{q3interesting}
This reduction presents a particularly interesting case. 
Here, we find the only non-linearisable mapping produced by the class of reductions considered in this paper, 
showing that indeed it is possible to achieve Painlev\'e type equations from Q3 using this method. 
Note that the constant $a$ in the reduction $\hat{u}=a/\ol {u}$ can be scaled away by a change of variables $u\mapsto \sqrt{a}\,u$. 
Hence we may assume without loss of generality $a=1$.
In this case, the autonomous reduction is given by
\begin{equation} \label{q3red}
 {\rm sinh}\,\alpha\,(\xa u^2\xub+1)-{\rm sinh}\,\beta\,(\xa \xub+u^2)-{\rm sinh}(\alpha-\beta)\,(\xa u+u \xub)=0,
\end{equation}
where $u=u(n)$, $\ol u=u(n+1)$ and $\underline{u}=u(n-1)$.
The degree growth of equation \eqref{q3red} is $1$, $2$, $5$, $8$, $13$, $18$, $25$, $32$,$\dots$, 
which appears to be quadratic, and thus not linearisable. 

It is known that singularity confinement can be used to find conditions on the parameters in an equation 
which permit deautonomisation whilst preserving integrability \cite{GramaniP:91}. 
With the benefit of hindsight, we replace the parameters $\alpha$ and $\beta$ with $\gamma$ and $z$ by
\begin{equation}\label{eqn:alphabeta_gammaz}
 \alpha=\gamma-z,\quad
 \beta=\gamma+z.
\end{equation}
Then, equation \eqref{q3red} can be rewritten as
\begin{equation} \label{q3na}
 {\rm sinh}(\gamma-z)\,(\xa u^2\xub+1)-{\rm sinh}(\gamma+z)\,(\xa \xub+u^2)+{\rm sinh}(2z)\,(\xa u+u \xub)=0.
\end{equation}
Surprisingly, this is precisely the change of parameters which was carried out in \cite{rcg:09} which led to deautonomisation. 
It would appear as though the difference and addition of the parameters \eqref{eqn:alphabeta_gammaz} 
naturally define the singular points in the reduction. 

To carry out singularity confinement, we ask when $\ol {u}$ is independent of $\underline{u}$. 
This happens when 
$(u,\ol {u})=(\pm {\rm e}^{\pm z},\pm {\rm e}^{\pm \gamma})$ (double sign in same order),
in which case equation \eqref{q3na} holds for any $\underline{u}$.
Moreover, from equation \eqref{q3na}$_{n\mapsto n+1}$ (i.e., equation \eqref{q3na} where $n\mapsto n+1$)
we obtain $\ol {\ol {u}}=\pm {\rm e}^{\pm z}$,
and then from equation \eqref{q3na}$_{n\mapsto n+2}$ we find that $\ol {\ol {\ol {u}}}$ is also free, 
so we have recovered the lost degree of freedom. 
%%%%%%%%%%%%%%%%%%%%%%%%%%%%%%%%%%%%%%%%%%%%%%%%%%%%%%%%%%%
%% 5.1 Deautonomisation
%%%%%%%%%%%%%%%%%%%%%%%%%%%%%%%%%%%%%%%%%%%%%%%%%%%%%%%%%%%
\subsection{Deautonomisation}
Let us consider the nonautonomous case of equation \eqref{q3na}
where $\gamma$ and $z$ both depend on the independent variable $n$,
that is, $\gamma=\gamma(n)$ and $z=z(n)$.
In the nonautonomous case, the solution enters the singularity through $u={\rm e}^z$, 
and leaves through $\ol {\ol {u}}={\rm e}^{\ol {\ol {z}}}$. 
Calculating $\ol {u}$ from equation \eqref{q3na} with $u={\rm e}^z$
and also from equation \eqref{q3na}$_{n\mapsto n+2}$ with $\ol {\ol {u}}={\rm e}^{\ol {\ol {z}}}$
in such a way that $\ul{u}$ and $\ol{\ol{\ol{u}}}$ are free, 
we obtain $\ol{u}={\rm e}^\gamma={\rm e}^{\ol{\ol{\gamma}}}$,
which means that $\gamma$ depends on whether $n$ is even or odd.
Let
\begin{equation}
 \gamma
 =\begin{cases}
 \gamma_e\quad (n=2k)\\
 \gamma_o\quad (n=2k+1)
 \end{cases}
\end{equation}
where $k\in\mathbb{Z}$ and $\gamma_e,\gamma_o\in\mathbb{C}$ are arbitrary constants.
Finally, we need to consider the consistency of the equation around $\ol {u}$. 
Substituting
\begin{equation}
 u={\rm e}^{z},\quad
 \ol {u}={\rm e}^{\gamma}, \quad
 \ol{\ol{u}}={\rm e}^{\ol{\ol{z}}},
\end{equation}
into equation \eqref{q3na}$_{n\mapsto n+1}$, we arrive at
\begin{equation}\label{zeqn}
 {\rm e}^{\ol{z}}({\rm e}^{2\ol{\gamma}}-{\rm e}^{2\ol{z}})({\rm e}^{z+\ol{\ol{z}}+2\gamma}+1)
 -{\rm e}^{\ol{z}}({\rm e}^{2\gamma}+{\rm e}^{z+\ol{\ol{z}}})({\rm e}^{2(\ol{z}+\ol{\gamma})}-1)
 +{\rm e}^{\gamma+\ol{\gamma}}({\rm e}^{4\ol{z}}-1)({\rm e}^{z}+{\rm e}^{\ol{\ol{z}}})=0.
\end{equation}
We can easily verify that
\begin{equation}
 z=(\gamma_e-\gamma_o)n+p,
\end{equation}
where $p\in\mathbb{C}$ is an arbitrary constant,
is a solution of equation \eqref{zeqn}.
Therefore, we obtain the nonautonomous form of equation \eqref{q3na}:
\begin{align} \label{nonautoq3red}
 &{\rm sinh}\left((\gamma_o-\gamma_e)n+\gamma-p\right)\,(\xa u^2\xub+1)
 -{\rm sinh}\left((\gamma_e-\gamma_o)n+\gamma+p\right)\,(\xa \xub+u^2)\notag\\
 &+{\rm sinh}\left(2(\gamma_e-\gamma_o)n+2p\right)\,(\xa u+\xa u)=0.
\end{align}
Moreover, letting
\begin{equation}
 x(k)=u(2k-1),\quad y(k)=u(2k),
\end{equation}
we recast equation \eqref{nonautoq3red} as the system 
\begin{subequations}\label{nonautoq3redsys}
\begin{align}
 &\widetilde{x}
 =-\cfrac{{\rm sinh}(z+\gamma_e)\,y^2-{\rm sinh}(2z)\,xy+{\rm sinh}(z-\gamma_e)}
 {{\rm sinh}(z-\gamma_e)\,xy^2-{\rm sinh}(2z)\,y+{\rm sinh}(z+\gamma_e)\,x}\,,\\
 &\widetilde{y}
 =-\cfrac{{\rm sinh}(z+\gamma_e)\,{\widetilde{x}\ }^2+{\rm sinh}(2(z+\gamma_e-\gamma_o))\,y\widetilde{x}+{\rm sinh}(z+\gamma_e-2\gamma_o)}
 {{\rm sinh}(z+\gamma_e-2\gamma_0)\,y{\widetilde{x}\ }^2-{\rm sinh}(2(z+\gamma_e-\gamma_o))\,\widetilde{x}+{\rm sinh}(z+\gamma_e)\,y}\,,
\end{align}
\end{subequations}
where $x=x(k)$, $y=y(k)$, $z=2(\gamma_e-\gamma_o)k+p$ and $~\widetilde{}~$ means $k\mapsto k+1$.
This system has base points at 
\begin{subequations}\label{basep}
\begin{align}
 &p_1:\,(x,y)=(-{\rm e}^{-z+\gamma_e-\gamma_o},-{\rm e}^{-\gamma_o}),
 &&p_2:\,(x,y)=({\rm e}^{-z+\gamma_e-\gamma_o},{\rm e}^{-\gamma_o}),\\
 &p_3:\,(x,y)=(-{\rm e}^{\gamma_e},-{\rm e}^z),
 &&p_4:\,(x,y)=({\rm e}^{\gamma_e},{\rm e}^z),\\
 &p_5:\,(x,y)=(-{\rm e}^{-\gamma_e},-{\rm e}^{-z}),
 &&p_6:\,(x,y)=({\rm e}^{-\gamma_e},{\rm e}^{-z}),\\
 &p_7:\,(x,y)=(-{\rm e}^{z-\gamma_e+\gamma_o},-{\rm e}^{\gamma_o}),
 &&p_8:\,(x,y)=({\rm e}^{z-\gamma_e+\gamma_o},{\rm e}^{\gamma_o}).
\end{align}
\end{subequations}
The base points $p_i$ $(i=1,\dots,4)$ and $p_i$ $(i=5,\dots,8)$ lie on
\begin{subequations}
\begin{align}
 &L_1:x={\rm e}^{-z+\gamma_e}y,\\
 &L_2:x={\rm e}^{z-\gamma_e}y,
\end{align}
\end{subequations}
respectively.
%%%%%%%%%%%%%%%%%%%%%%%%%%%%%%%%%%%%%%%%%%%%%%%%%%%%%%%%%%%
%% 5.2 The space of initial condition and the affine Weyl group
%%%%%%%%%%%%%%%%%%%%%%%%%%%%%%%%%%%%%%%%%%%%%%%%%%%%%%%%%%%
\subsection{The space of initial condition and the affine Weyl group}
The relations between rational surfaces and affine Weyl groups are studied in \cite{Looijenga:81}.
Furthermore, in \cite{sak:01} Sakai relates them to the discrete Painlev\'e systems.
In this section, following Sakai's method, we describe the time evolution of system \eqref{nonautoq3redsys} in terms of the corresponding affine Weyl group.

Let $\pi: X \to \mathbb{P}^1\times\mathbb{P}^1$ denote blow up of $\mathbb{P}^1\times\mathbb{P}^1$ 
at the eight base points given in \eqref{basep}.
The Picard group of rational surface $X$ is given by
\begin{equation}
 {\rm Pic}(X)=\mathbb{Z}h_x+\mathbb{Z}h_y+\sum_{i=1}^8\mathbb{Z}e_i,
\end{equation}
which is defined earlier in Section \ref{section:initialvaluespace}.
The anticanonical divisor of rational surface $X$:
\begin{equation}
 \delta=2h_x+2h_y-\sum_{i=1}^8 e_i,
\end{equation}
is uniquely decomposed into the prime divisors as the following:
\begin{equation}
\delta=D_1+D_2,
\end{equation}
where $D_1$ and $D_2$ are the proper transforms of the lines $L_1$ and $L_2$:
\begin{equation}
 D_1=h_x+h_y-e_1-e_2-e_3-e_4,\quad
 D_2=h_x+h_y-e_5-e_6-e_7-e_8,
\end{equation}
respectively.
Therefore, the rational surface $X$ is said to be of the $A_1^{(1)}$ type, following Sakai's classification \cite{sak:01}.\\

The orthogonal compliment $\delta^\bot=\{v\in {\rm Pic}(X)\,|\,(v|D_1)=(v|D_2)=0\}$ has eight generators
\begin{align}
 &\alpha_1=-e_1+e_2,\quad
 \alpha_2=-e_2+e_3,\quad
 \alpha_3=-e_3+e_4,\quad
 \alpha_4=h_y-e_4-e_5,\\
 &\alpha_5=e_5-e_6,\quad
 \alpha_6=e_6-e_7,\quad
 \alpha_7=e_7-e_8,\quad
 \alpha_0=h_x-h_y,
\end{align}
which satisfy
\begin{equation}
 \delta=\alpha_1+2\alpha_2+3\alpha_3+4\alpha_4+3\alpha_5+2\alpha_6+\alpha_7+2\alpha_0.
\end{equation}
The root lattice $Q(E_7^{(1)})=\sum_{i=0}^7\mathbb{Z}\alpha_i$ can be identified as type $E_7^{(1)}$ since
\begin{equation}
 (\alpha_i|\alpha_j)=-C_{ij}
\end{equation}
where $(C_{ij})_{i,j=0}^7$ is the Cartan matrix of type $E_7^{(1)}$:
\begin{equation}
 (C_{ij})_{i,j=0}^7
 =\begin{pmatrix}
 2&0&0&0&-1&0&0&0\\
 0&2&-1&0&0&0&0&0\\
 0&-1&2&-1&0&0&0&0\\
 0&0&-1&2&-1&0&0&0\\
 -1&0&0&-1&2&-1&0&0\\
 0&0&0&0&-1&2&-1&0\\
 0&0&0&0&0&-1&2&-1\\
 0&0&0&0&0&0&-1&2
 \end{pmatrix}.
\end{equation}
The corresponding Dynkin diagram is given by the following.
%%%%%%%%%%%%%%%%%%%%%%%%
%% Figure 5.1
%%%%%%%%%%%%%%%%%%%%%%%%
\begin{figure}[ht!] 
\centering
\begin{tikzpicture}[scale=0.5]
%%% draw the lines
\draw (0,0) -- (2,0);
\draw (2,0) -- (4,0);
\draw (4,0) -- (6,0);
\draw (6,0) -- (8,0);
\draw (8,0) -- (10,0);
\draw (10,0) -- (12,0);
\draw (6,0) -- (6,2);
%%% lables the simple roots
\filldraw [black] (0,0) node [anchor=north] {$\al_1$}  circle(1ex);
\filldraw [black] (2,0) node [anchor=north] {$\al_2$}  circle(1ex);
\filldraw [black] (4,0) node [anchor=north] {$\al_3$}  circle(1ex);
\filldraw [black] (6,0) node [anchor=north] {$\al_4$}  circle(1ex);
\filldraw [black] (8,0) node [anchor=north] {$\al_5$}  circle(1ex);
\filldraw [black] (10,0) node [anchor=north] {$\al_6$}  circle(1ex);
\filldraw [black] (12,0) node [anchor=north] {$\al_7$}  circle(1ex);
\filldraw [black] (6,2) node [anchor=south] {$\al_0$}  circle(1ex);
\end{tikzpicture}
\end{figure}
%%%%%%%%%%%%%%%%%%%%%%%%
%%%%%%%%%%%%%%%%%%%%%%%%

We define the reflections $s_i$ $(i=0,\dots,7)$ across the hyperplane orthogonal to the root $\alpha_i$ by
\begin{equation}
 s_i(v)=v+(v|\alpha_i)\alpha_i 
\end{equation}
for all $v\in {\rm Pic}(X)$ and the diagram automorphism $\sigma$ by
\begin{equation}
 \sigma:
 (e_1,e_2,e_3,e_4,e_5,e_6,e_7,e_8)
 \mapsto(e_8,e_7,e_6,e_5,e_4,e_3,e_2,e_1).
\end{equation}
The actions of $\widetilde{W}(E_7^{(1)})=\langle s_0,\dots,s_7,\sigma\rangle$ on Pic$(X)$ and $Q(E_7^{(1)})$ are
summarised in Tables \ref{table:picard} and \ref{table:root}, respectively.
Blanks in the tables represent invariance.
Moreover, $\widetilde{W}(E_7^{(1)})$ forms the extended affine Weyl group of type $E_7^{(1)}$. 
Namely, the generators satisfy the fundamental relations
\begin{align}
 {s_i}^2&=1,\quad i=0,\dots,7,\\
 (s_i\circ s_{i+1})^3&=1,\quad i=1,\dots,6,\\
 (s_4\circ s_0)^3&=1,\\
 (s_i\circ s_j)^2&=1,\quad\text{otherwise},\\
 \sigma^2&=1,\\
 \sigma\circ s_{(1,2,3,5,6,7)}&=s_{(7,6,5,3,2,1)}\circ \sigma.
\end{align}

%%%%%%%%%%%%%%%%%%%%%%%%%%%%%%%
%% Table 5.1
%%%%%%%%%%%%%%%%%%%%%%%%%%%%%%%
{\small
\begin{table}[htb]
$$\begin{array}{|l||l|l|l|l|l|l|l|l|l|}
\hline
&s_1&s_2&s_3&s_4&s_5&s_6&s_7&s_0&\sigma\\
\hline\hline
h_x&&&&h_x+h_y-e_4-e_5&&&&h_y&\\
\hline
h_y&&&&&&&&h_x&\\
\hline
e_1&e_2&&&&&&&&e_8\\
\hline
e_2&e_1&e_3&&&&&&&e_7\\
\hline
e_3&&e_2&e_4&&&&&&e_6\\
\hline
e_4&&&e_3&h_y-e_5&&&&&e_5\\
\hline
e_5&&&&h_y-e_4&e_6&&&&e_4\\
\hline
e_6&&&&&e_5&e_7&&&e_3\\
\hline
e_7&&&&&&e_6&e_8&&e_2\\
\hline
e_8&&&&&&&e_7&&e_1\\
\hline
\end{array}$$
\caption{The action of $\widetilde{W}(E_7^{(1)})$ on Pic$(X)$.}
\label{table:picard}
\end{table}
}
%%%%%%%%%%%%%%%%%%%%%%%%%%%%%%%
%% Table 5.2
%%%%%%%%%%%%%%%%%%%%%%%%%%%%%%%
{\small
\begin{table}[htb]
$$\begin{array}{|l||l|l|l|l|l|l|l|l|l|}
\hline
&s_1&s_2&s_3&s_4&s_5&s_6&s_7&s_0&\sigma\\
\hline\hline
\alpha_1&-\alpha_1&\alpha_1+\alpha_2&&&&&&&\alpha_7\\
\hline
\alpha_2&\alpha_1+\alpha_2&-\alpha_2&\alpha_2+\alpha_3&&&&&&\alpha_6\\
\hline
\alpha_3&&\alpha_2+\alpha_3&-\alpha_3&\alpha_3+\alpha_4&&&&&\alpha_5\\
\hline
\alpha_4&&&\alpha_3+\alpha_4&-\alpha_4&\alpha_4+\alpha_5&&&\alpha_4+\alpha_0&\\
\hline
\alpha_5&&&&\alpha_4+\alpha_5&-\alpha_5&\alpha_5+\alpha_6&&&\alpha_3\\
\hline
\alpha_6&&&&&\alpha_5+\alpha_6&-\alpha_6&\alpha_6+\alpha_7&&\alpha_2\\
\hline
\alpha_7&&&&&&\alpha_6+\alpha_7&-\alpha_7&&\alpha_1\\
\hline
\alpha_0&&&&\alpha_4+\alpha_0&&&&-\alpha_0&\\
\hline
\end{array}$$
\caption{The action of $\widetilde{W}(E_7^{(1)})$ on $Q(E_7^{(1)})$.}
\label{table:root}
\end{table}
}
%%%%%%%%%%%%%%%%%%%%%%%%%%%%%%%
%%%%%%%%%%%%%%%%%%%%%%%%%%%%%%%

Define the translations $T_i={R_i}^2$ $(i=1,\dots,7,0)$ where
\begin{align}
 &R_1=s_2\circ s_1\circ s_2\circ s_3\circ s_4\circ s_5\circ s_6\circ s_0\circ s_4\circ s_3\circ s_5\circ s_4\notag\\
 &\hspace{3em}\circ s_0\circ s_7\circ s_6\circ s_5\circ s_4\circ s_3\circ s_2,\\
 &R_2=s_1\circ s_2\circ R_1\circ s_2\circ s_1,\\
 &R_3=s_2\circ s_3\circ R_2\circ s_3\circ s_2,\\
 &R_4=s_3\circ s_4\circ R_3\circ s_4\circ s_3,\\
 &R_5=\sigma \circ R_3\circ \sigma,\\
 &R_6=\sigma \circ R_2\circ \sigma,\\
 &R_7=\sigma \circ R_1\circ \sigma,\\
 &R_0=s_4\circ s_0\circ R_4\circ s_0\circ s_4.
\end{align} 
The actions of the translations on the generators of $Q(E_7^{(1)})$ are given by
\begin{align}
 &T_1:\bm{\alpha}
 \mapsto 
 \bm{\alpha}+(2\delta,-\delta,0,0,0,0,0,0),\\
 &T_2:\bm{\alpha}
 \mapsto
 \bm{\alpha}+(-\delta,2\delta,-\delta,0,0,0,0,0),\\
 &T_3:\bm{\alpha}
 \mapsto
 \bm{\alpha}+(0,-\delta,2\delta,-\delta,0,0,0,0),\\
 &T_4:\bm{\alpha}
 \mapsto
 \bm{\alpha}+(0,0,-\delta,2\delta,-\delta,0,0,-\delta),\\
 &T_5:\bm{\alpha}
 \mapsto
 \bm{\alpha}+(0,0,0,-\delta,2\delta,-\delta,0,0),\\
 &T_6:\bm{\alpha}
 \mapsto
 \bm{\alpha}+(0,0,0,0,-\delta,2\delta,-\delta,0),\\
 &T_7:\bm{\alpha}
 \mapsto
 \bm{\alpha}+(0,0,0,0,0,-\delta,2\delta,0),\\
 &T_0:\bm{\alpha}
 \mapsto
 \bm{\alpha}+(0,0,0,-\delta,0,0,0,2\delta),
\end{align}
where $\bm{\alpha}=(\alpha_1,\dots,\alpha_7,\alpha_0)$.
Note that $T_i$ $(i=1,\dots,7,0)$ commute with each other and 
\begin{equation}
 T_1\circ {T_2}^2\circ {T_3}^3\circ {T_4}^4\circ {T_5}^3\circ {T_6}^2\circ T_7\circ {T_0}^2=1.
\end{equation}

Let the mapping $\varphi :(x,y)\mapsto (\widetilde {x},\widetilde {y})$ be the time evolution of the system \eqref{nonautoq3redsys}.
In a similar manner as Section \ref{q1xbisxhsec},
we obtain the action of the mapping $\varphi$ on Pic$(X)$ as
\begin{equation}
 \varphi:
 \begin{pmatrix}
 h_x\\h_y\\e_1\\e_2\\e_3\\e_4\\e_5\\e_6\\e_7\\e_8
 \end{pmatrix}
 \mapsto
 \begin{pmatrix}
 1&2&0&0&-1&-1&-1&-1&0&0\\
 2&5&-1&-1&-2&-2&-2&-2&-1&-1\\
 1&2&-1&0&-1&-1&-1&-1&0&0\\
 1&2&0&-1&-1&-1&-1&-1&0&0\\
 0&1&0&0&-1&0&0&0&0&0\\
 0&1&0&0&0&-1&0&0&0&0\\
 0&1&0&0&0&0&-1&0&0&0\\
 0&1&0&0&0&0&0&-1&0&0\\
 1&2&0&0&-1&-1&-1&-1&-1&0\\
 1&2&0&0&-1&-1&-1&-1&0&-1
 \end{pmatrix}
 \begin{pmatrix}
 h_x\\h_y\\e_1\\e_2\\e_3\\e_4\\e_5\\e_6\\e_7\\e_8
 \end{pmatrix}.
\end{equation}
We can verify that the mapping $\varphi$ can be written in terms of the generators of $\widetilde{W}(E_7^{(1)})$ as
\begin{align}\label{eqn:phi_si}
 \varphi
 =\sigma&\circ  s_3\circ s_5\circ 
 s_4\circ s_5\circ s_0\circ s_3\circ 
 s_4\circ s_0\circ s_6\circ s_5\circ s_2\circ s_3\circ s_1\circ s_2\circ s_7\circ s_4\notag\\
 &\circ s_3\circ s_2\circ s_5\circ s_6\circ
 s_4\circ s_5\circ s_3\circ 
 s_4\circ s_3\circ s_7\circ s_6\circ s_7\circ s_1\circ s_2\circ s_1\circ s_0.
\end{align}
The action of the mapping $\varphi$ is given by
\begin{equation}
 \varphi:~
 \begin{pmatrix}
 \alpha_1\\ \alpha_2\\ \alpha_3\\ \alpha_4\\ \alpha_5\\ \alpha_6\\ \alpha_7\\ \alpha_0
 \end{pmatrix}
 \mapsto
 \begin{pmatrix}
 -1&0&0&0&0&0&0&0\\
 0&-1&-1&-2&-1&0&0&-1\\
 0&0&-1&0&0&0&0&0\\
 1&2&4&5&4&2&1&2\\
 0&0&0&0&-1&0&0&0\\
 0&0&-1&-2&-1&-1&0&-1\\
 0&0&0&0&0&0&-1&0\\
 -1&-2&-3&-4&-3&-2&-1&-1
 \end{pmatrix}
 \begin{pmatrix}
 \alpha_1\\ \alpha_2\\ \alpha_3\\ \alpha_4\\ \alpha_5\\ \alpha_6\\ \alpha_7\\ \alpha_0
 \end{pmatrix},
\end{equation}
which implies that $\varphi$ is not a translation on the root lattice,
but the action of two successive iterations of
$\varphi$ is given by
\begin{equation}
 \varphi^2:~
 \begin{pmatrix}
 \alpha_1\\ \alpha_2\\ \alpha_3\\ \alpha_4\\ \alpha_5\\ \alpha_6\\ \alpha_7\\ \alpha_0
 \end{pmatrix}
 \mapsto
 \begin{pmatrix}
 1&0&0&0&0&0&0&0\\
 0&1-\delta&0&0&0&0&0&0\\
 0&0&1&0&0&0&0&0\\
 0&0&0&1+2\delta&0&0&0&0\\
 0&0&0&0&1&0&0&0\\
 0&0&0&0&0&1-\delta&0&0\\
 0&0&0&0&0&0&1&0\\
 0&0&0&0&0&0&0&1-2\delta
 \end{pmatrix}
 \begin{pmatrix}
 \alpha_1\\ \alpha_2\\ \alpha_3\\ \alpha_4\\ \alpha_5\\ \alpha_6\\ \alpha_7\\ \alpha_0
 \end{pmatrix},
\end{equation}
which means $\varphi^2$ can be expressed in terms of translations $T_3\circ {T_4}^2\circ T_5$.
%%%%%%%%%%%%%%%%%%%%%%%%%%%%%%%%%%%%%%%%%%%%%%%%%%%%%%%%%%%
%% 5.3 Reconstruction of system \eqref{nonautoq3redsys} from the birational representation of $\widetilde{W}(E_7^{(1)})$
%%%%%%%%%%%%%%%%%%%%%%%%%%%%%%%%%%%%%%%%%%%%%%%%%%%%%%%%%%%
\subsection{Reconstruction of system \eqref{nonautoq3redsys} from the birational representation of $\widetilde{W}(E_7^{(1)})$}
Since the configuration of the base points of system \eqref{nonautoq3redsys} does not change through the deautonomisation, 
we use the birational representation of $\widetilde{W}(E_7^{(1)})$,
which may be verified directly,
 given in \cite{ahjn:13} where the geometrical properties of an autonomous version of system \eqref{nonautoq3redsys} were investigated.

The $q$-Painlev\'e system of $A_1^{(1)}$-surface type has the following eight base points \cite{MSY:03}:
\begin{subequations}\label{eqn:appendix_bps}
\begin{align}
 &p_1: (x,y)=(-vu_1,-u_1), 
 &&p_2: (x,y)=(-vu_2,-u_2),\\
 &p_3: (x,y)=(-vu_3,-u_3),
 &&p_4: (x,y)=(-vu_4,-u_4),\\
 &p_5: (x,y)=(-v^{-1}u_5,-u_5),
 &&p_6: (x,y)=(-v^{-1}u_6,-u_6),\\
 &p_7: (x,y)=(-v^{-1}u_7,-u_7),
 &&p_8: (x,y)=(-v^{-1}u_8,-u_8).
\end{align}
\end{subequations}
The above base points coincide with base points \eqref{basep}
when the parameters are specialized by setting 
\begin{subequations}\label{eqns:special_para}
\begin{align}
 &{u_1}^{1/2}={\rm e}^{-\gamma_o/2},\quad
 {u_2}^{1/2}=-{\rm i}{\rm e}^{-\gamma_o/2},\quad
 {u_3}^{1/2}={\rm e}^{z/2},\quad
 {u_4}^{1/2}=-{\rm i}{\rm e}^{z/2},\\
 &{u_5}^{1/2}={\rm e}^{-z/2},\quad
 {u_6}^{1/2}=-{\rm i}{\rm e}^{-z/2},\quad
 {u_7}^{1/2}={\rm e}^{\gamma_o/2},\quad
 {u_8}^{1/2}=-{\rm i}{\rm e}^{\gamma_o/2},\\
 &v={\rm e}^{-z+\gamma_e}.
\end{align}
\end{subequations}
The birational actions of $\widetilde{W}(E_7^{(1)})$ on the parameters $v$ and $u_j$ are as follows  \cite{ahjn:13}:
\begin{align*}
 s_i:&(u_i,u_{i+1})\mapsto (u_{i+1},u_i),\quad i=1,2,3,5,6,7,\\
 s_4:&(u_4,u_5,v)\mapsto(u_5,u_4,v{u_4}^{1/2}{u_5}^{-1/2}),\\
 s_0:&(u_1,u_2,u_3,u_4,u_5,u_6,u_7,u_8,v)\\
 &\mapsto(vu_1,vu_2,vu_3,vu_4,v^{-1}u_5,v^{-1}u_6,v^{-1}u_7,v^{-1}u_8,v^{-1}),\\
 \sigma:&(u_1,u_2,u_3,u_4,u_5,u_6,u_7,u_8)
 \mapsto({u_8}^{-1},{u_7}^{-1},{u_6}^{-1},{u_5}^{-1},{u_4}^{-1},{u_3}^{-1},{u_2}^{-1},{u_1}^{-1}),
\end{align*}
and those on the variables $x$ and $y$ are given by
\begin{align*}
 &s_4(x)
 ={u_4}^{1/2}{u_5}^{1/2}
 \cfrac{(v-v^{-1})xy+(vu_4-v^{-1}u_5)x+(-u_4+u_5)y}{(-u_4+u_5)x+(vu_4-v^{-1}u_5)y+(v-v^{-1})u_4u_5},\\
 &s_0:(x,y)\mapsto (y,x),\\
 &\sigma:(x,y)\mapsto(x^{-1},y^{-1}).
\end{align*}

Now we are in the position to give the explicit expressions of $q$-\Pa equations of $A_1^{(1)}$-surface type.
First, we consider the translation $T_0$.
The translation $T_0$ act on the parameters as
\begin{align}
 T_0:~&
 (u_1,u_2,u_3,u_4)
 \mapsto
 (q^{-1}u_1,q^{-1}u_2,q^{-1}u_3,q^{-1}u_4),\\
 &(u_5,u_6,u_7,u_8,v)
 \mapsto
 (qu_5,qu_6,qu_7,qu_8,q^2v),
\end{align}
where
\begin{equation}
 q=\cfrac{v^2{u_1}^{1/2}{u_2}^{1/2}{u_3}^{1/2}{u_4}^{1/2}}{{u_5}^{1/2}{u_6}^{1/2}{u_7}^{1/2}{u_8}^{1/2}},
\end{equation}
and on the variables as
\begin{subequations}\label{eqns:qPA1_T0}
\begin{align}
 &\cfrac{(fT_0(g)-q^2v^2)(fg-v^2)}{(fT_0(g)-1)(fg-1)}
 =\cfrac{(f-b_1 v)(f-b_2 v)(f-b_3 v)(f-b_4 v)}{(f-b_5)(f-b_6)(f-b_7)(f-b_8)},\\
 &\cfrac{(fg-v^2)({T_0}^{-1}(f)g-q^{-2}v^2)}{(fg-1)({T_0}^{-1}(f)g-1)}
 =\cfrac{(g-{b_1}^{-1} v)(g-{b_2}^{-1} v)(g-{b_3}^{-1} v)(g-{b_4}^{-1} v)}{(g-{b_5}^{-1})(g-{b_6}^{-1})(g-{b_7}^{-1})(g-{b_8}^{-1})},
\end{align}
\end{subequations}
where
\begin{equation}
 f=-v^{1/2}x,\quad g=-\cfrac{v^{1/2}}{y},\quad b_i=v^{1/2}u_i\quad (i=1,\dots,4),\quad b_j=\cfrac{u_j}{v^{1/2}}\quad (j=5,\dots,8).
\end{equation}
The system \eqref{eqns:qPA1_T0} is usually referred to as a $q$-Painlev\'e equation of $A_1^{(1)}$-surface type\cite{MSY:03,sak:01}.

Next, we consider the mapping $\varphi$.
From \eqref{eqn:phi_si} the action of $\varphi$ on the parameters $v$ and $u_i$ is given by
\begin{align}
 \varphi:~&(u_1,u_2,u_3)\mapsto 
 \left(
 \cfrac{{u_2}^{1/2}}{{u_1}^{1/2}{u_7}^{1/2}{u_8}^{1/2}}\,,
 \cfrac{{u_1}^{1/2}}{{u_2}^{1/2}{u_7}^{1/2}{u_8}^{1/2}}\,,
 \cfrac{v^2{u_1}^{1/2}{u_2}^{1/2}u_4}{u_5u_6{u_7}^{1/2}{u_8}^{1/2}}
 \right),\\
 &(u_4,u_5,u_6)\mapsto 
 \left(
 \cfrac{v^2{u_1}^{1/2}{u_2}^{1/2}u_3}{u_5u_6{u_7}^{1/2}{u_8}^{1/2}}\,,
 \cfrac{u_6{u_7}^{1/2}{u_8}^{1/2}}{v^2{u_1}^{1/2}{u_2}^{1/2}u_3u_4}\,,
 \cfrac{u_5{u_7}^{1/2}{u_8}^{1/2}}{v^2{u_1}^{1/2}{u_2}^{1/2}u_3u_4}
 \right),\\
 &(u_7,u_8,v)~\,\mapsto 
 \left(
 \cfrac{{u_8}^{1/2}}{{u_1}^{1/2}{u_2}^{1/2}{u_7}^{1/2}}\,,
 \cfrac{{u_7}^{1/2}}{{u_1}^{1/2}{u_2}^{1/2}{u_8}^{1/2}}\,,
 \cfrac{{u_5}^{1/2}{u_6}^{1/2}{u_7}^{1/2}{u_8}^{1/2}}{v{u_1}^{1/2}{u_2}^{1/2}{u_3}^{1/2}{u_4}^{1/2}}
 \right),
\end{align}
and that on the variables $x$ and $y$ is given by
\begin{subequations}\label{eqns:phi_fg}
\begin{align}
 &\varphi(x)
 =\cfrac{1}{{u_3}^{1/2}{u_4}^{1/2}{u_5}^{1/2}{u_6}^{1/2}}\,
 \cfrac{A_1y^2+A_2xy+A_3+A_4x+A_5y}{A_6xy^2+A_7y+A_8x+A_9y^2+A_{10}xy},\\
 &\varphi(y)
 =\cfrac{1}{{u_1}^{1/2}{u_2}^{1/2}{u_7}^{1/2}{u_8}^{1/2}}\,
 \cfrac{B_1\varphi(x)^2+B_2y\varphi(x)+B_3+B_4y\varphi(x)^2+B_5\varphi(x)}{B_6y\varphi(x)^2+B_7\varphi(x)+B_8y+B_9y\varphi(x)+B_{10}},
\end{align}
\end{subequations}
where
{\allowdisplaybreaks
\begin{align}
 &A_1=v^2u_3u_4-u_5u_6,
 &&A_2=v(-u_3u_4+u_5u_6),\\
 &A_3=(-1+v^2)u_3u_4u_5u_6,
 &&A_4=v(-u_3u_4(u_5+u_6)+(u_3+u_4)u_5u_6),\\
 &A_5=v^2u_3u_4(u_5+u_6)-(u_3+u_4)u_5u_6,
 &&A_6=-1+v^2,\\
 &A_7=v(-u_3u_4+u_5u_6),
 &&A_8=v^2u_3u_4-u_5u_6,\\
 &A_9=v(-(u_3+u_4)+(u_5+u_6)),
 &&A_{10}=v^2(u_3+u_4)-(u_5+u_6),
 \end{align}
{\allowdisplaybreaks
\begin{align}
 &B_1=vu_1u_2{u_3}^{1/2}{u_4}^{1/2}{u_5}^{1/2}{u_6}^{1/2}u_7u_8(v^2u_3u_4-u_5u_6),\\
 &B_2=v^4u_1u_2{u_3}^2{u_4}^2-{u_5}^2{u_6}^2u_7u_8,\\
 &B_3=v{u_3}^{1/2}{u_4}^{1/2}{u_5}^{1/2}{u_6}^{1/2}(u_5u_6u_7u_8-v^2u_1u_2u_3u_4),\\
 &B_4=v{u_3}^{1/2}{u_4}^{1/2}{u_5}^{1/2}{u_6}^{1/2}(v^2u_1u_2u_3u_4(u_7+u_8)-(u_1+u_2)u_5u_6u_7u_8),\\
 &B_5=v^2u_3u_4u_5u_6((u_1+u_2)u_7u_8-u_1u_2(u_7+u_8)),\\
 &B_6=v{u_3}^{1/2}{u_4}^{1/2}{u_5}^{1/2}{u_6}^{1/2}(u_5u_6u_7u_8-v^2u_1u_2u_3u_4),\\
 &B_7=v^4u_1u_2{u_3}^2{u_4}^2-{u_5}^2{u_6}^2u_7u_8,\\
 &B_8=v{u_3}^{1/2}{u_4}^{1/2}{u_5}^{1/2}{u_6}^{1/2}(v^2u_3u_4-u_5u_6),\\
 &B_9=v^2u_3u_4u_5u_6(-(u_1+u_2)+(u_7+u_8)),\\
 &B_{10}=v{u_3}^{1/2}{u_4}^{1/2}{u_5}^{1/2}{u_6}^{1/2}(v^2(u_1+u_2)u_3u_4-u_5u_6(u_7+u_8)).
\end{align}
}

After we make the specialization \eqref{eqns:special_para},
the action of $\varphi$ is reduced to
\begin{subequations}\label{eqns:phi_fg_special}
\begin{align}
 &\varphi:(z,\gamma_e,\gamma_o)\mapsto(z+2(\gamma_e-\gamma_o),\gamma_e,\gamma_o),\notag\\
 &\varphi(x)
 =-\cfrac{{\rm sinh}(z+\gamma_e)\,y^2-{\rm sinh}(2z)\,xy+{\rm sinh}(z-\gamma_e)}
 {{\rm sinh}(z-\gamma_e)\,xy^2-{\rm sinh}(2z)\,y+{\rm sinh}(z+\gamma_e)\,x}\,,\\
 &\varphi(y)
 =-\cfrac{{\rm sinh}(z+\gamma_e)\,\varphi(x)^2+{\rm sinh}(2(z+\gamma_e-\gamma_o))\,y\varphi(x)+{\rm sinh}(z+\gamma_e-2\gamma_o)}
 {{\rm sinh}(z+\gamma_e-2\gamma_0)\,y\varphi(x)^2-{\rm sinh}(2(z+\gamma_e-\gamma_o))\,\varphi(x)+{\rm sinh}(z+\gamma_e)\,y}\,,
\end{align}
\end{subequations}
which is equivalent to system \eqref{nonautoq3redsys} with the correspondence $\varphi=\widetilde{}$~.

In what follows, a discrete Painlev\'e equation being ``full-parameter'' means that
it has the same number of parameters as the corresponding surface-type discrete Painlev\'e equation in Sakai's list \cite{sak:01}.
For instance, system \eqref{eqns:phi_fg} is the full-parameter version of system \eqref{eqns:phi_fg_special} (or system \eqref{nonautoq3redsys}).
Often a discrete Painlev\'e equation can be extended up to the full-parameter version  by singularity confinement \cite{GramaniP:91,RGH:91}.
However, system \eqref{nonautoq3redsys} has a reduced number of parameters after applying the singularity confinement criterion.
The reason is some of the coefficients in system \eqref{eqns:phi_fg} become zero:
\begin{equation}
 A_4=A_5=A_9=A_{10}=B_4=B_5=B_9=B_{10}=0,
\end{equation} 
under the condition \eqref{eqns:special_para} .
In such a case, singularity confinement cannot be used to extend the equation up to the full-parameter version.
%%%%%%%%%%%%%%%%%%%%%%%%%%%%%%%%%%%%%%%%%%%%%%%%%%%%%%%%%%%
%%%%%%%%%%%%%%%%%%%%%%%%%%%%%%%%%%%%%%%%%%%%%%%%%%%%%%%%%%%

\section{Conclusion}\label{qredconc}
This paper has shown that M\"obius type reductions of Q1, Q2 and Q3 give rise to second order mappings, all but one of which
are linearisable. In the  linearisable cases the explicit linearisation was given using a constructive methodology that requires no guess work.
The only mapping that was not linearisable was produced by a reduction from Q3. Singularity confinement was applied to this mapping
to deautonomise the parameters while preserving integrability. 
The nonautonomous form of the equation was shown to be a $q$-Painlev\'e equation, in the sense of \cite{sak:01}. This is the first case of an equation with the rational surface of $A_1^{(1)}$ type being found as a reduction of a lattice equation. The advantage of utilising the geometry of the space of initial conditions is that it enables a coordinate invariant description of the equation, that is, the identification of the equation without the need to find a transformation to some known example. Future work should carry out the analysis of Q4 from the ABS list, as well as exploring higher order staircase reductions of the ABS equations as a natural generalisation of the approach used in this paper \cite{KNT:11, MSY:03, RGH:91}.

\noindent{\bf Acknowledgement.} 
This work has been supported by DFG Research Unit "Polyhedral Surfaces" and
the Australian Research Council grant DP130100967.

\end{document}